\documentclass[%
 reprint,
superscriptaddress,
nofootinbib,
 amsmath,amssymb,
 aps,
prx,
]{revtex4-2}
\usepackage[dvipsnames]{xcolor}
\usepackage{graphicx}
\usepackage{dcolumn}
\usepackage{bm}
\usepackage{hyperref}

\usepackage{braket}
\usepackage[utf8]{inputenc} 
\usepackage{mathtools}
\usepackage[bottom]{footmisc}
\usepackage{hyperref}
\usepackage{amsfonts}
\usepackage{amsmath}
\usepackage{slashed}
\usepackage{amsthm}
\usepackage[normalem]{ulem}
\usepackage{soul}

\begin{document}

\preprint{APS/123-QED}

\title{Feature-energy duality of topological boundary states in multilayer quantum spin Hall insulator}

\author{Yueh-Ting Yao}
\affiliation{Department of Physics,\;National Cheng Kung University,\;Tainan,\;70101,\;Taiwan}

\author{Xiaoting Zhou}
\affiliation{Department of Physics,\;Northeastern\;University,\;Boston,\;Massachusetts,\;02115,\;USA}

\author{Yi-Chun Hung}
\affiliation{Department of Physics,\;Northeastern\;University,\;Boston,\;Massachusetts,\;02115,\;USA}

\author{Hsin Lin}
\affiliation{Institute of Physics,\;Academia Sinica,\;Taipei,\;115201,\;Taiwan}

\author{Arun Bansil}
\affiliation{Department of Physics,\;Northeastern\;University,\;Boston,\;Massachusetts,\;02115,\;USA}

\author{Tay-Rong Chang}
\email{u32trc00@phys.ncku.edu.tw}
\affiliation{Department of Physics,\;National Cheng Kung University,\;Tainan,\;70101,\;Taiwan}
\affiliation{Center for Quantum Frontiers of Research and Technology (QFort),\;Tainan,\;70101,\;Taiwan}
\affiliation{Physics Division, National Center for Theoretical Sciences,\;Taipei,\;10617,\;Taiwan}

\date{\today} 

\begin{abstract}
Gapless topological boundary states characterize nontrivial topological phases arising from the bulk-boundary correspondence in symmetry-protected topological materials, such as the emergence of helical edge states in a two-dimensional $\mathbb{Z}_2$ topological insulator. However, the incorporation of symmetry-breaking perturbation terms in the Hamiltonian leads to the gapping of these edge bands, resulting in missing these crucial topological boundary states. In this work, we systematically investigate the robustness of bulk-boundary correspondence in the quantum spin Hall insulator via recently introduced feature spectrum topology. Our findings present a comprehensive understanding of feature-energy duality, illustrating that the aggregate number of gapless edge states in the energy-momentum (\textit{E-k}) map and the non-trivial edge states in the $\hat{S}_z$ feature spectrum equals the spin Chern number of multilayer quantum spin Hall insulator. We identify a van der Waals material bismuth bromide $\rm(Bi_4Br_4)$ as a promising candidate through first-principles calculations. Our work not only unravels the intricacies of bulk-boundary correspondence but also charts a course for exploring quantum spin Hall insulators with high spin-Chern number.
\end{abstract}

\maketitle




\section{\label{sec:level1}Introduction}

\par The two-dimensional (2D) quantum spin Hall effect has attracted significant attention over the past two decades due to the unconventional spin propagation on the edge of the materials and its potential application in spintronics devices as well as quantum computations \cite{NO1,NO2,NO3,NO4,NO5}. Quantum spin Hall insulator (QSHI), characterized by spin Chern number (SChN, $C_s$), are considered prime candidates for realizing the quantum spin Hall effect in condensed matter materials. The term QSHI has been utilized to describe the $\mathbb{Z}_2$ class of the topological insulator in 2D, attributed to the existence of gapless topological nontrivial helical edge states resulting from the bulk-boundary correspondence \cite{NO6,NO7}. However, this understanding is somewhat misleading. The $\mathbb{Z}_2$ index is ill-defined in bulk band structures if the original gapless edge states open a gap in the absence of time-reversal symmetry \cite{NO7,NO8,NO9}. In contrast, the SChN can be defined in the thermodynamic limit and generalized the Chern number without time-reversal symmetry \cite{NO10}. Consequently, the time-reversal breaking quantum spin Hall effect emerges as a form of QSHI classified by SChN \cite{NO4,NO8,NO9,NO10}. Following the well-established concept of bulk-boundary correspondence in non-trivial topological phases, we anticipate the appearance of a gapless topological edge state on the boundary of QSHI without time-reversal symmetry. However, research on bulk-boundary correspondence under symmetry-breaking conditions of QSHI has remained elusive for many years. In addition to time-reversal breaking in monolayer QSHI, bilayer QSHI with each layer carrying SChN $C_s=1$, exhibit $\mathbb{Z}_2=0$ and $C_s=2$ in the entire system. In this scenario, interlayer coupling hybridizes the gapless edge states of each layer. Since the electron spins are not generally conserved, such hybridization results in gapped edge state in bilayer QSHI with $C_s=2$ , challenging the concept of bulk boundary correspondence \cite{NO11}. These two cases underscore the invalidity of the bulk-boundary correspondence in QSHI when the original gapless edge states are gapped by a perturbation.

\par In this work, we investigate the bulk-boundary correspondence via feature spectrum topology \cite{NO10,NO12,NO13,NO14} with $\hat{S}_z$ feature operator. We first introduce a perturbation involving a Zeeman exchange filed in monolayer Kane-Mele model \cite{NO6,NO7,NO15}. Our calculations reveal the expected gapped edge state of QSHI in the \textit{E-k} map under time-reversal symmetry-breaking perturbation. Concurrently, a non-trivial edge state emerges, connecting the spin-up and down sectors separated by a gap in the $\hat{S}_z$ feature spectrum. Moving on to the bilayer Kane-Mele model \cite{NO16,NO17,NO18} with QSHI phase possessing $C_s=2$, we observe an energy gap in the edge states within the \textit{E-k} map due to interlayer coupling. In contrast, a pair of non-trivial edge states manifest within the gap of the bulk feature spectrum, which equilavent to the SChN in the bilayer Kane-Mele model. To investigate the feature spectrum topology in a realistic material, we study the feature eigenvalues, topological invariants, and corresponding edge states in a van der Waals material $\rm{Bi_4Br_4}$, employing first-principles calculations. Our results demonstrate a feature-energy duality, revealing that the total number of gapless edge states in the \textit{E-k} map and non-trivial edge states in the feature spectrum equals the SChN in multilayer $\rm{Bi_4Br_4}$. Furthermore, we note an increase in the SChN with stacking layers, demonstrating the presence of a QSHI insulating phase with a high-spin Chern number in $\rm{Bi_4Br_4}$.

\section{\label{sec:level1}Feature Spectrum Topology in Quantum Spin Hall Insulator}

\par Very recently, an alternative perspective to understanding the essence of topological phases of matter, namely the feature spectrum topology, has been proposed \cite{NO12,NO13}. This approach involves selecting a particular quantum operator $\hat{O}$ through which the projection operator $(P)$ projects the occupied Bloch state wave functions onto a quantum operator $\hat{O}$, forming the feature operator $\hat{\mathbb{F}}_O=P\hat{O}P$. This operation partitions the wavefunction of occupied states into distinct sectors. The expectation value of $\hat{\mathbb{F}}_O$ is termed the feature spectrum $O_{mk}$, serving as an alternative representation of band structure, which replaces the Hamiltonian acting on Bloch states $\hat{H}\left | \psi_{nk} \right \rangle=E_{nk}\left | \psi_{nk} \right \rangle$ by the feature operator $\hat{\mathbb{F}}_O\left | \widetilde{\psi}_{mk} \right \rangle=O_{mk}\left | \widetilde{\psi}_{mk} \right \rangle$, where $\left | \widetilde{\psi}_{mk} \right \rangle$ denotes the eigenstates of $\hat{\mathbb{F}}_O$ at each momentum vector $k$ and index of sector $m$. Similar to the Chern number in Chern insulator \cite{NO19,NO20}, the corresponding topological invariants $C_{\hat{O}}$ in the feature spectrum represent the Chern number defined by particular feature sectors associated with the feature operator $\hat{\mathbb{F}}_O$ in $\mathbb{Z}$ class \cite{NO10,NO12,NO13,NO14}. Moreover, the universality of the feature operator leads to a well-defined feature spectrum even when the system lacks the symmetry corresponding to the selected quantum operator.

\par In examining the feature spectrum topology of QSHI in the Kane-Mele model, we select $\hat{S}_z$ as the quantum operator to formulate the feature operator $\hat{\mathbb{F}}_{S_z}=P\hat{S}_zP$. The four-band Kane-Mele tight-binding Hamiltonian \cite{NO6,NO7} is expressed as

\begin{equation}\label{eq:01}
\begin{split}
    H_0=&\  t \sum_{\left \langle i,j \right \rangle_{\alpha}} c_{i\alpha}^{\dag}c_{j\alpha} + \frac{i\lambda_I}{3\sqrt{3}}\sum_{\left \langle \left \langle i,j \right \rangle \right \rangle_{\alpha\beta}} \nu_{ij}c_{i\alpha}^{\dag}s_zc_{j\beta}\\
    & - i\frac{2}{3}\lambda_{so}\sum_{\left \langle \left \langle i,j \right \rangle \right \rangle_{\alpha\beta}}\mu_ic_{i\alpha}^{\dag}(\boldsymbol{s\times\hat{d}_{ij}})^z_{\alpha\beta} c_{j\beta}
\end{split}
\end{equation}

\noindent where $c_{i\alpha}^{\dag}$ and $c_{j\alpha}$ represent the creation and the annihilation operator of an electron with spin polarization $\alpha$ at sites $i$ and $j$, respectively. The symbols $\left \langle i,j \right \rangle$ and $\left \langle \left \langle i,j \right \rangle \right \rangle$ run over the nearest-neighbor and next-nearest-neighbor hopping sites. $\boldsymbol{s}=(s_x,s_y,s_z)$ denotes the Pauli matrix for the spin degree of freedom. The first term represents the nearest-neighbor hopping with hopping strength $t$. The second term denotes intrinsic spin-orbital coupling (SOC), in which $\lambda_I$ is the strength of intrinsic SOC and $\nu_{ij}=1(-1)$ indicates clockwise(counter-clockwise) next-nearest-neighbor hopping with respect to positive $\hat{z}$ axis. The third term introduces next-nearest-neighbor Rashba SOC with strength $\lambda_{so}$. Here $\mu_i=1(-1)$ denotes A(B) site and $\boldsymbol{\hat{d}}_{ij}$ is the unit vector of the next-nearest-neighbor hopping. The last term breaks $SU(2)$ symmetry \cite{NO6} and couples the spin-up and spin-down states, resulting in a non-$\hat{S}_z$-conserved Hamiltonian \cite{NO12}. 

\par
\begin{figure}[ht]
\centering
\includegraphics[width=\linewidth]{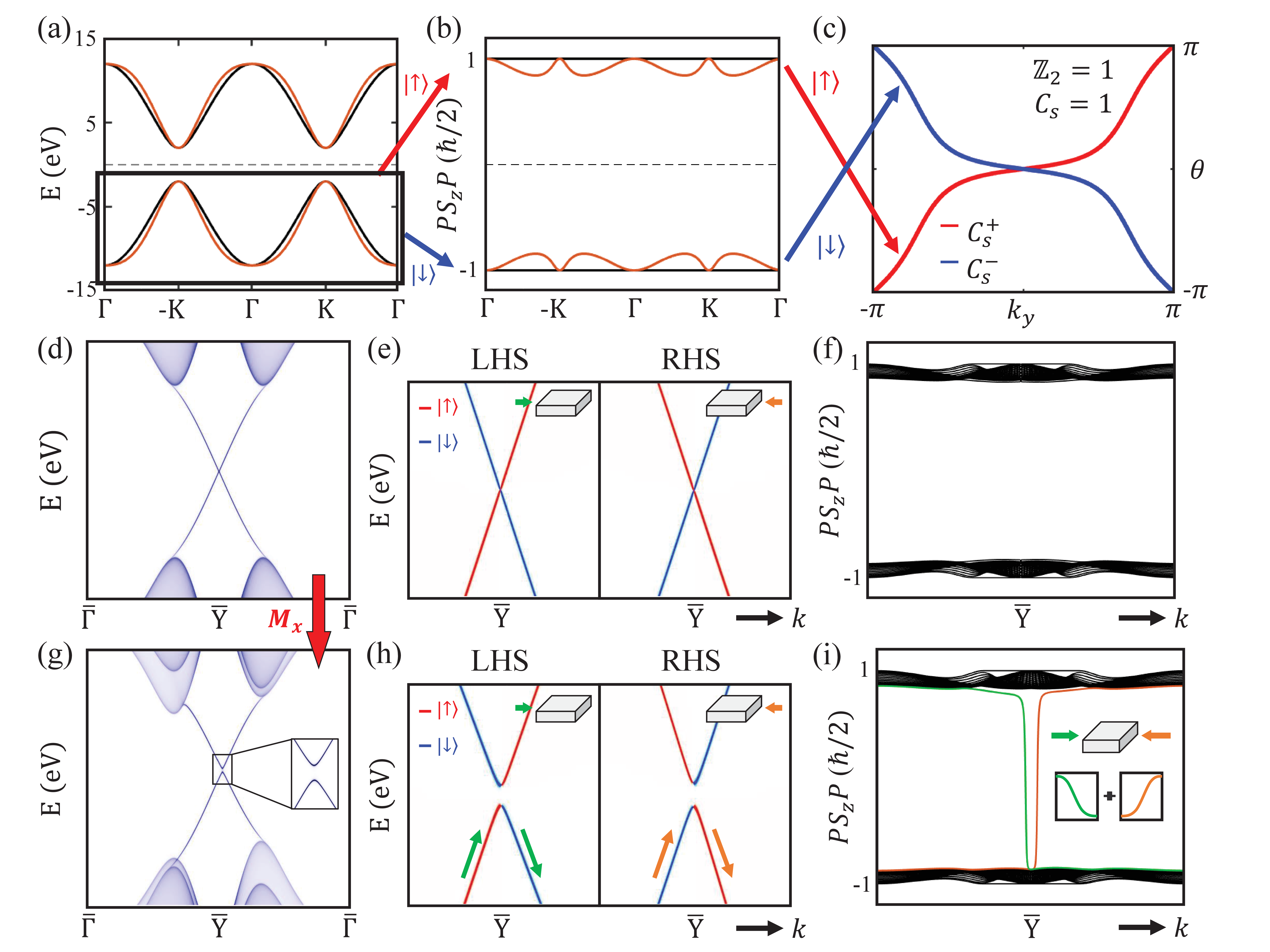}
\caption{
\textbf{The monolayer Kane-Mele model with and without time-reversal symmetry.} 
(a)Black line: band structure with $t=4$, $\lambda=-2$, and $\lambda_{so}=0$. Brown line: band structure with $t=4$, $\lambda=-2$, and $\lambda_{so}=2$. (b)$\left \langle PS_zP \right \rangle$ feature spectrum of $\hat{S}_z$-conserved (black line) and non-$\hat{S}_z$-conserved model (brown line). (c)The spin-resolve Wilson’s loop indicates the $C_s=1$ and $\mathbb{Z}_2=1$. The red and blue line indicates the $C_s^+$ and $C_s^-$, respectively. (d)The edge states with time-reversal symmetry in the \textit{E-k} map. (e)Spin-polarized edge states with time-reversal symmetry in the \textit{E-k} map. The red and blue lines denote spin-up and spin-down, respectively. The green and orange arrows in the inset indicate the edge states from the LHS boundary and RHS boundary, respectively. (f)$\hat{S}_z$ feature edge states in the Kane-Mele model with time-reversal symmetry. (g) Same as (d), but time-reversal symmetry breaking. (h) Same as (e), but time-reversal symmetry breaking. The green and orange arrows indicate the direction of $\hat{S}_z$ propagation on the edge states from LHS boundary and RHS boundary, respectively. The green and orange arrows in the inset indicate the edge states from the LHS boundary and RHS boundary, respectively. (i) Same as (f), but time-reversal symmetry breaking. The green and orange line denotes the nontrivial edge states in the edge feature spectrum from LHS boundary and RHS boundary, respectively. The green and orange arrows in the inset indicate the edge states from the LHS boundary and RHS boundary, respectively.
}
\label{fig:01}
\end{figure}

 For simplicity, we initially focus on the first two terms. The resulting band structure is illustrated in Fig.~\ref{fig:01}(a) with black lines. The gap of the band structure is induced by the strength of intrinsic SOC $\lambda_I$, leading to a nontrivial topological phase of QSHI. Each Bloch state wavefunction $\left | \psi_{nk}^s \right \rangle$ with opposite spins is doubly degenerate due to the preservation of both time-reversal symmetry and spatial-inversion symmetry in the Hamiltonian. Here $n$ is the band index, $k$ is the momentum vector in the Brillouin zone (BZ), and $s$ is the spin index for the $\hat{S}_z$ spin-up and spin-down states. As the Rashba SOC is absent, the Hamiltonian can be decomposed into two spin-diagonal blocks. Based on the $\hat{S}_z$ characteristic equation $\hat{S}_z\left | \psi_{nk}^s \right \rangle=\pm \frac{\hbar}{2}\left | \psi_{nk}^s \right \rangle$, each Bloch state of the $\hat{S}_z$-conserved Hamiltonian can be categorized into $\pm \frac{\hbar}{2}$ eigenvalues in $\hat{S}_z$ quantum number. Notably, the eigenvalues of feature operator $\hat{\mathbb{F}}_{S_z}$ reveal two distinct flat bands divided into $\pm \frac{\hbar}{2}$ two sectors (black lines in Fig.~\ref{fig:01}(b)). This non-dispersive feature band structure indicates the conservation of $\hat{S}_z$ in the Hamiltonian.

\par The calculation of the SChN for the QSHI within the framework of feature spectrum topology can be determined based on the eigenstates of each sector in the feature spectrum \cite{NO6,NO9,NO22,NO23,NO24,NO25}. We compute the Chern number for each spin $C_s^{\pm}$ by using the well-established method, spin-resolve Wilson’s loop \cite{NO10,NO12,NO13,NO14,NO21}. The calculation result is shown in Fig.~\ref{fig:01}(c). The trajectories of the Wannier center charge for spin-up and spin-down sectors display opposite open curves traversing the entire BZ once, indicating the $C_s^{+}=1$ and $C_s^{-}=-1$. The total SChN is defined by $C_s=(C_s^+-C_s^-)/2$. Fig.1~\ref{fig:01}(c) shows the $C_s=1$, indicating the presence of a QSHI. It is well-known that the QSHI is associated with a topological insulator (TI) in 2D classified by the $\mathbb{Z}_2$ index \cite{NO7,NO22}, where $\mathbb{Z}_2=(C_s,\rm{mod\ 2})$. The calculated $\mathbb{Z}_2=1$ in this case confirms a  $\mathbb{Z}_2$ TI.

\par The brown lines in Fig.~\ref{fig:01}(a) show the band structure when the Rashba SOC is introduced. The feature spectrum of $\hat{\mathbb{F}}_{S_z}$ is no longer exhibits +1 and -1 while displaying dispersive eigenvalues due to the coupling between spin-up and spin-down induced by Rashba SOC (brown lines in Fig.~\ref{fig:01}(b)). The eigenstates of the feature spectrum in this situation consist of the linear combination of the occupied states of Bloch wavefunctions with opposite spins. Although $\hat{S}_z$ is no longer conserved in the Hamiltonian, it is noteworthy that the sectors of the feature eigenvalues remain separated by a gap. Consequently, the SChN is well-defined in this non-$\hat{S}_z$-conserved system \cite{NO8,NO10,NO14}. Our calculation shows $C_s$ and $\mathbb{Z}_2$ both equal 1.

\subsection{\label{sec:level2}Feature spectrum topology in QSHI without time-reversal symmetry}

\par The Kane-Mele model with time-reversal breaking \cite{NO15} can be described as

\begin{equation}\label{eq:02}
\begin{split}
    H_M=\ H_0+ M\sum_{i\alpha} c_{i\alpha}^{\dag}s_zc_{i\alpha} + l\sum_{i\alpha}\mu_{i} c_{i\alpha}^{\dag}c_{i\alpha}
\end{split}
\end{equation}

\noindent The first term corresponds to the original Kane-Mele model in Eq.~\ref{eq:01}. The second term introduces the Zeeman exchange field $M$ to break the time-reversal symmetry. The third term presents the sublattice onsite energy with strength $l$, which eliminates the out-of-plane mirror symmetry $\mathcal{M}_z$. 

\par The topological edge band structure from Eq.~\ref{eq:01} is shown in Fig.~\ref{fig:01}(d). The gapless Dirac point at the $\bar{Y}$ point is protected by time-reversal symmetry. The projection of the expectation value of the spin operator onto the topological edge Dirac state exhibits opposite spin texture on the left-hand-side (LHS) boundary and the right-hand-side (RHS) boundary, as illustrated in Fig.~\ref{fig:01}(e). Upon introducing the $\rm{2^{nd}}$ and $\rm{3^{rd}}$ terms in Eq.~\ref{eq:02} as perturbations in the calculation, the topological Dirac band structure opens a gap at the $\bar{Y}$ point due to time-reversal symmetry breaking (Fig.~\ref{fig:01}(g)). Similar to the effect of Rashba SOC, the gap-opening terms in Eq.~\ref{eq:02} perturb the dispersion relation in the feature eigenvalues. Therefore, the SChN can still be calculated based on the eigenstates of each spin sector in the feature band structure of $\hat{\mathbb{F}}_{S_z}$ generated by the time-reversal symmetry-breaking Hamiltonian. Our calculation shows $C_s=1$, indicating a QSHI with time-reversal symmetry-breaking perturbation \cite{NO8,NO9,NO10}. According to the bulk-boundary correspondence, a non-zero $\mathbb{Z}_2$ class topological invariant number will induce a corresponding gapless boundary state at the system boundary. However, this one-to-one correspondence is invalid in topological edge band structure without the protection of time-reversal symmetry in QSHI. To better comprehend the nature of bulk-boundary correspondence, we calculate the feature spectrum of a finite-sized ribbon structure with zigzag edges (Fig.~\ref{fig:01}(i)). We observe two non-trivial edge states from the LHS boundary and the RHS boundary (labeled in green and orange lines, respectively) manifesting within the gap of the bulk feature eigenvalues. The edge state from the LHS boundary is connected from the spin-up sector to the spin-down sector in the feature subspace, akin to the band dispersion of the non-trivial edge states in the typical Chern insulator in the \textit{E-k} map \cite{NO20}. The connectivity of the non-trivial edge state in the feature spectrum can be understood from the evolution of the spin texture on the gapped topological edge state in the \textit{E-k} map. With $\bar{Y}$ point as the reference center, the $\hat{S}_z$ spin-up state flips to the $\hat{S}_z$ spin-down state when $\hat{S}_z$ propagate from left to right along the $k$ direction on the edge band structure (green arrows in Fig.~\ref{fig:01}(h)). This spin-flipping in the gapped topological edge band structure corresponds to the flow of the spin sector on the edge feature spectrum (green line in Fig.~\ref{fig:01}(i)). The spin texture in the RHS boundary exhibits the opposite behavior to the LHS (orange arrows in Fig.~\ref{fig:01}(h)), leading to reverse connectivity of edge states in the feature spectrum (orange line in Fig.~\ref{fig:01}(i)). In contrast to the situation of time-reversal symmetry breaking, the gapless topological Dirac state results in an ill-defined edge feature spectrum at the $\bar{Y}$ point, which leads to the edge feature spectrum remaining in an insulating state (Fig.~\ref{fig:01}(f)). This phenomenon, gaps opening/closing of the topological edge state in the \textit{E-k} map, corresponds to the closing/opening in the feature spectrum on the system boundary, termed the \textit{feature-energy duality} \cite{NO12}. The feature-energy duality demonstrates the robustness of bulk-boundary correspondence with the non-trivial topology classified by the feature Chern number $C_{\hat{O}}$.

\subsection{\label{sec:level2}Feature spectrum topology in bilayer QSHI with interlayer coupling}

\par A single layer of QSHI carries SChN $C_s=1$, it is therefore intuitively inferred that increasing the number of stacked layers of QSHI will elevate the Chern number and the corresponding topological edge states carried by the system if the interaction between layers can be treated as perturbations. To examine this conjecture, we introduce the bilayer Kane-Mele model with interlayer coupling \cite{NO16,NO17,NO18}, expressed as

\begin{equation}\label{eq:03}
\begin{split}
    H_{Bilayer}=&\ H_0+ t_l\sum_{il\alpha}\mu_i c_{il\alpha}^{\dag}c_{i(l+1)\alpha} \\
    &+ t_o\sum_{il\alpha}\xi_lc_{il\alpha}^{\dag}c_{il\alpha}
\end{split}
\end{equation}

\noindent The first term is the monolayer Kane-Mele model from Eq.~\ref{eq:01}. The second term represents the interlayer coupling, where $c_{il\alpha}^{\dag}$ and $c_{il\alpha}$ denotes the creation and the annihilation operator of an electron with spin polarization $\alpha$ at site $i$ in layer $l$, respectively. $\mu_i=1(-1)$ corresponds to A(B) site with strength $t_l$, breaking the in-plane mirror symmetry $\mathcal{M}_{//}$ . The third term is the interlayer onsite energy, where $\xi_l=1(-1)$ for the first(second) layer, which eliminates out-of-plane mirror symmetry $\mathcal{M}_z$ with strength $t_o$.

\par
\begin{figure}[ht]
\centering
\includegraphics[width=\linewidth]{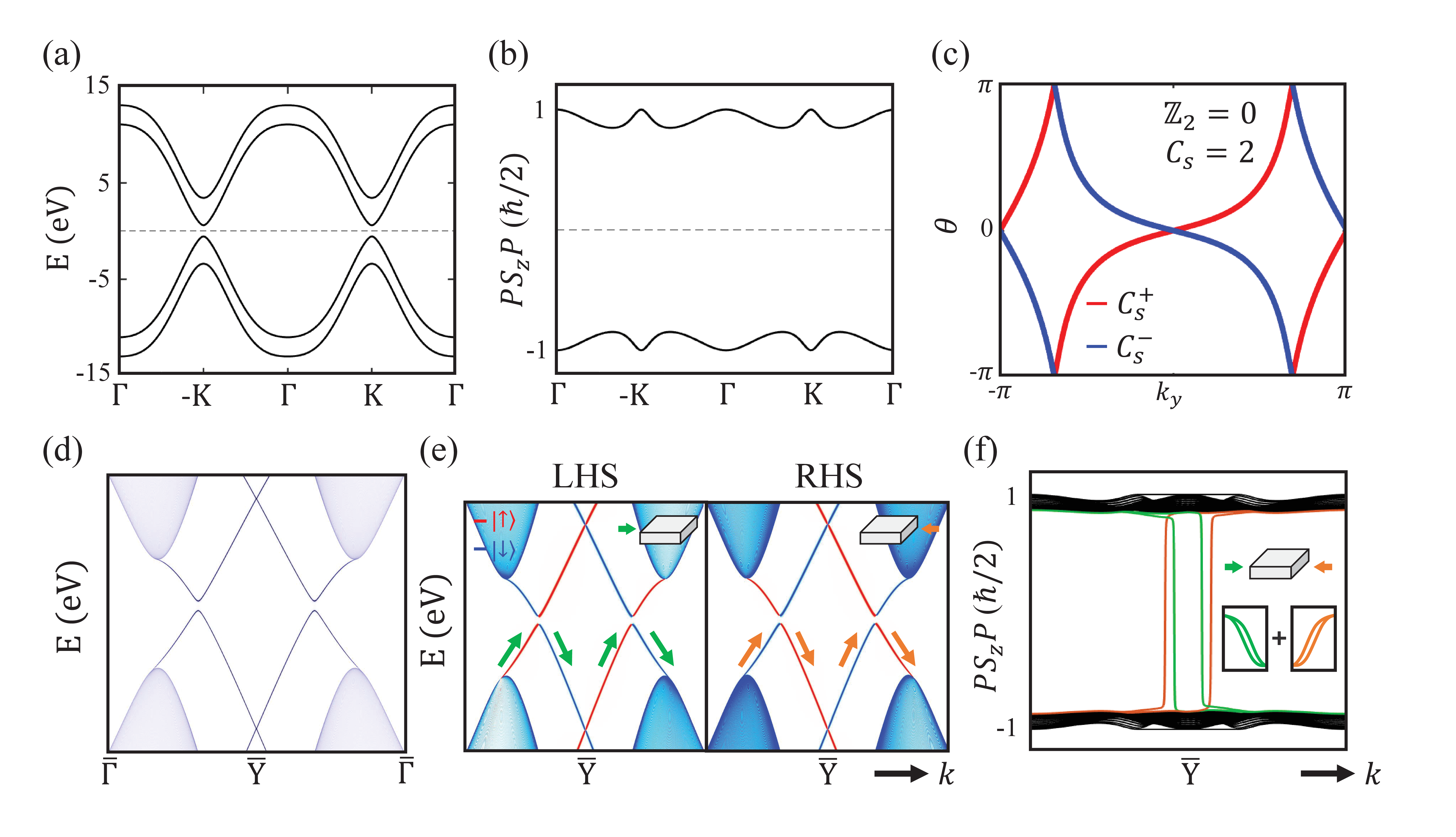}
\caption{\textbf{The bilayer Kane-Mele model with interlayer coupling.} (a)The band structure with $t_l=t_o=1$. (b)$\left \langle PS_zP \right \rangle$ feature spectrum in bilayer Kane-Mele model with interlayer coupling. The feature spectrum is doubly degenerate in layer degree of freedom. (c)The spin-resolve Wilson’s loop indicates the $C_s=2$ and $\mathbb{Z}_2=0$. The red and blue line indicates the $C_s^+$ and $C_s^-$, respectively. (d)The edge states with interlayer coupling in the \textit{E-k} map. (e)Spin-polarized edge states with interlayer coupling in the \textit{E-k} map. The red and blue lines denote spin-up and spin-down, respectively. The green and orange arrows indicate the direction of $\hat{S}_z$ propagation on the edge states from the LHS boundary and RHS boundary, respectively. (f)$\hat{S}_z$ feature edge states in the bilayer Kane-Mele model with interlayer coupling. The green and orange lines denote the nontrivial edge states in the edge feature spectrum from the LHS boundary and RHS boundary, respectively. The green and orange arrows in the inset indicate the edge states from the LHS boundary and RHS boundary, respectively.
}
\label{fig:02}
\end{figure}

 The band structure of bilayer Kane-Mele model is shown in Fig.~\ref{fig:02}(a). The band gap and band splitting result from the intralayer SOC and interlayer coupling, respectively. Fig.~\ref{fig:02}(d) presents the edge state band structure, revealing two pairs of edge states from the top and bottom layers emerging in the bulk gap. These edge bands cross each other at generic $k$ point in the BZ. Since the crossing points lack symmetry protection, the interlayer coupling disrupts these gapless points, creating gaps in the edge bands. This outcome indicates the topological invariant number $\mathbb{Z}_2=0$, signifying the bilayer Kane-Mele model as a $\mathbb{Z}_2=0$ trivial insulator. The $\hat{S}_z$ feature spectrum of bilayer Kane-Mele model is presented in Fig.~\ref{fig:02}(b), showing dispersive feature eigenvalues caused by the intralayer Rashba SOC. Since the parameters of SOC within each layer are identical, the feature spectrum exhibits doubly degeneracy throughout the whole BZ. We calculate the SChN based on the eigenstates of each sector in the feature spectrum. The trajectories of the Wannier center charge for spin-up and spin-down sectors display opposite open curves traversing the entire BZ twice, indicating $C_s=2$ in the bilayer Kane-Mele model (Fig.~\ref{fig:02}(c)). The non-zero SChN corresponds to the existence of a non-trivial edge state emerging in the feature edge spectrum in accordance with the bulk-boundary correspondence. In Fig.~\ref{fig:02}(f), we observe two pairs of non-trivial edge states from the LHS and RHS boundaries (labeled in green and orange lines, respectively) manifest within the gap of the bulk feature eigenvalues. The connectivity of the non-trivial edge states in the feature spectrum (green line in Fig.~\ref{fig:02}(f)) can be interpreted as the spin flip of $\hat{S}_z$ propagating in two pairs of gapped topological edge bands (green arrows in Fig.~\ref{fig:02}(e)). Our results clearly reveal the robustness of bulk-boundary correspondence in the feature-energy duality: the gapped edge states of energy bands (Fig.~\ref{fig:02}(e)) correspond to the non-trivial feature edge states (Fig.~\ref{fig:02}(f)). Our calculation shows that the $\mathbb{Z}_2=0$ in the bilayer QSHI, while $C_s=2$, suggesting that multilayer QSHI is allowed to carry high-spin Chern numbers.

\section{\label{sec:level1}Feature spectrum topology in van der Waals material $\bf{Bi_4Br_4}$ }

\par The feature spectrum topology provides an advanced perspective for comprehending the bulk-boundary correspondence. Since it depends only on valence bands within the band structure, it is well-suited for the evaluation in first-principles calculations, allowing for quantitative predictions in condensed matter materials. Here, we consider $\rm{Bi_4Br_4}$, which is a van der Waals material predicted to be a 2D topological insulator with $\mathbb{Z}_2=1$ in monolayer \cite{NO11,NO23}. In the following, we demonstrate that multilayer $\rm{Bi_4Br_4}$ exemplifies the bulk-boundary correspondence of the feature spectrum topology in the Kane-Mele models we discussed earlier.

\par
\begin{figure}[ht]
\centering
\includegraphics[width=\linewidth]{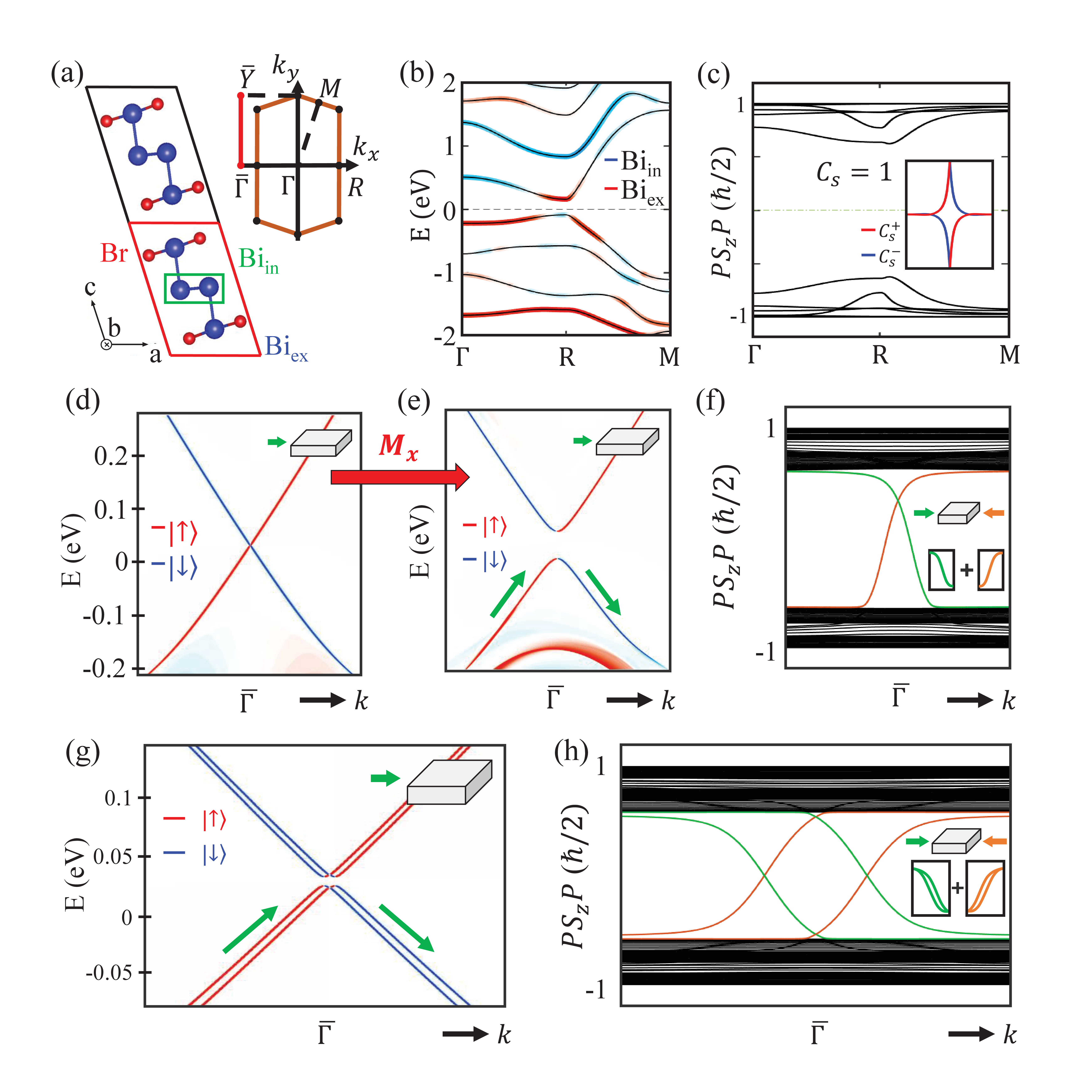}
\caption{\textbf{The monolayer and bilayer $\bf{Bi_4Br_4}$} (a)The crystal structure of $\beta$-$\rm{Bi_4Br_4}$. The inset is a 2D Brillouin zone (BZ). The red square indicates the monolayer structure, and the black square is for the bilayer structure. (b)The band structure of monolayer $\rm{Bi_4Br_4}$ with atomic decomposition to Bi atoms. (c)$\left \langle PS_zP \right \rangle$ feature spectrum of monolayer $\rm{Bi_4Br_4}$. The inset is calculated spin-resolved Wilson’s loop, indicating $C_s=1$ and $\mathbb{Z}_2=1$. The red and blue line indicates the $C_s^+$ and $C_s^-$, respectively. (d)Spin-polarized edge states of monolayer $\rm{Bi_4Br_4}$ with time-reversal symmetry in the \textit{E-k} map. The green arrow in the inset indicates the edge states from the LHS boundary. (e)Same as (d), but time-reversal symmetry breaking. The green arrows indicate the direction of $\hat{S}_z$ propagation on the edge states from the LHS boundary. The green arrow in the inset indicates the edge states from the LHS boundary. (f)The $\left \langle PS_zP \right \rangle$ edge feature spectrum of monolayer $\rm{Bi_4Br_4}$ time-reversal symmetry breaking. The green and orange lines denote the nontrivial edge states in the edge feature spectrum from the LHS boundary and RHS boundary, respectively. The green and orange arrows in the inset indicate the edge states from the LHS boundary and RHS boundary, respectively. (g) Spin-polarized edge states of bilayer $\rm{Bi_4Br_4}$ with interlayer coupling in the \textit{E-k} map. The green arrows indicate the direction of $\hat{S}_z$ propagation on the edge states from the LHS boundary. The green arrow in the inset denotes the edge states from the LHS boundary. (h) The $\left \langle PS_zP \right \rangle$ edge feature spectrum of bilayer $\rm{Bi_4Br_4}$ interlayer coupling. The green and orange lines denote the nontrivial edge states in the edge feature spectrum from the LHS boundary and RHS boundary, respectively.
}
\label{fig:03}
\end{figure}

 The  $\beta$-$\rm{Bi_4Br_4}$ crystallizes in the monoclinic space group $C_{2h}^3 (C2/m)$ with AA stacking between alternating layers (Fig.~\ref{fig:03}(a)). Fig.~\ref{fig:03}(b) presents the band structure of monolayer $\rm{Bi_4Br_4}$, decomposed into atomic orbitals of Bi atoms, via first-principles calculations (See method). The band inversion between external Bi ($\rm{Bi_{ext}}$) and internal Bi ($\rm{Bi_{in}}$) atoms around the Fermi level at $R$ point in the BZ originates from the strong spin-orbital coupling. The gapless topological edge state in Fig.~\ref{fig:03}(d) reflects the non-trivial bulk-boundary correspondence in monolayer $\rm{Bi_4Br_4}$, which is consistent with previous calculations \cite{NO11,NO23}. Considering a time-reversal symmetry breaking perturbation, an in-plane Zeeman exchange field $M_x$, this gapless edge state opens a gap and transforms into an insulating phase in the \textit{E-k} map (Fig.~\ref{fig:03}(e)). The $\hat{S}_z$ feature spectrum of monolayer $\rm{Bi_4Br_4}$ is shown in Fig.~\ref{fig:03}(c). The dispersive feature eigenvalues result from the complicated non-$\hat{S}_z$-conserved SOC in real material (e.g. Rashba SOC). The SChN of monolayer $\rm{Bi_4Br_4}$ is calculated using the method of spin-resolve Wilson’s loop in each spin sector of gapped feature spectrum. The result $C_s=1$ indicates a QSHI phase (inset of Fig.~\ref{fig:03}(c)), corresponding to the well-konwn $\mathbb{Z}_2=1$ in monolayer $\rm{Bi_4Br_4}$ \cite{NO11,NO23,NO24,NO25}. The SChN $C_s=1$ remains in our calculation even with the introduction of $M_x$ as a perturbation, indicating monolayer $\rm{Bi_4Br_4}$ is a time-reversal symmetry breaking QSHI. The edge feature spectrum with the 1D ribbon structure of monolayer $\rm{Bi_4Br_4}$ shows two non-trivial edge states in the bulk gap of feature spectrum (Fig.3~\ref{fig:03}(f)). This feature edge state flows in the feature spectrum (green line in Fig.~\ref{fig:03}(f)), corresponding the characteristics of the $\hat{S}_z$ spin filp on the gapped edge band structure in the \textit{E-k} map (green arrows in Fig.~\ref{fig:03}(e)), as introduced in the monolayer Kane-Mele model with time-reversal symmetry breaking perturbation. 

\par In bilayer $\rm{Bi_4Br_4}$ (AA stacking of monolayer), the originally paired gapless edge bands of each layer give rise to an insulator state due to non-negligible interlayer coupling (Fig.~\ref{fig:03}(g)), showing a $\mathbb{Z}_2=0$ trivial phase. By contrast, the feature spectrum topology analysis provides a non-trivial SChN $C_s=2$. This is because each $\rm{Bi_4Br_4}$ layer undergoes a band inversion, such that the SChN of each layer is $C_s=1$, resulting in total $C_s=2$. We observe that two pairs of non-trivial edge states corresponding to the $C_s=2$ in the edge feature spectrum flow from the $\hat{S}_z$ spin-up sector to spin-down sector along the $k$-direction (green lines in Fig.~\ref{fig:03}(h)). Our result reveals that bulk-boundary correspondence is still valid in bilayer $\rm{Bi_4Br_4}$.

\begin{figure}[ht]
\centering
\includegraphics[width=\linewidth]{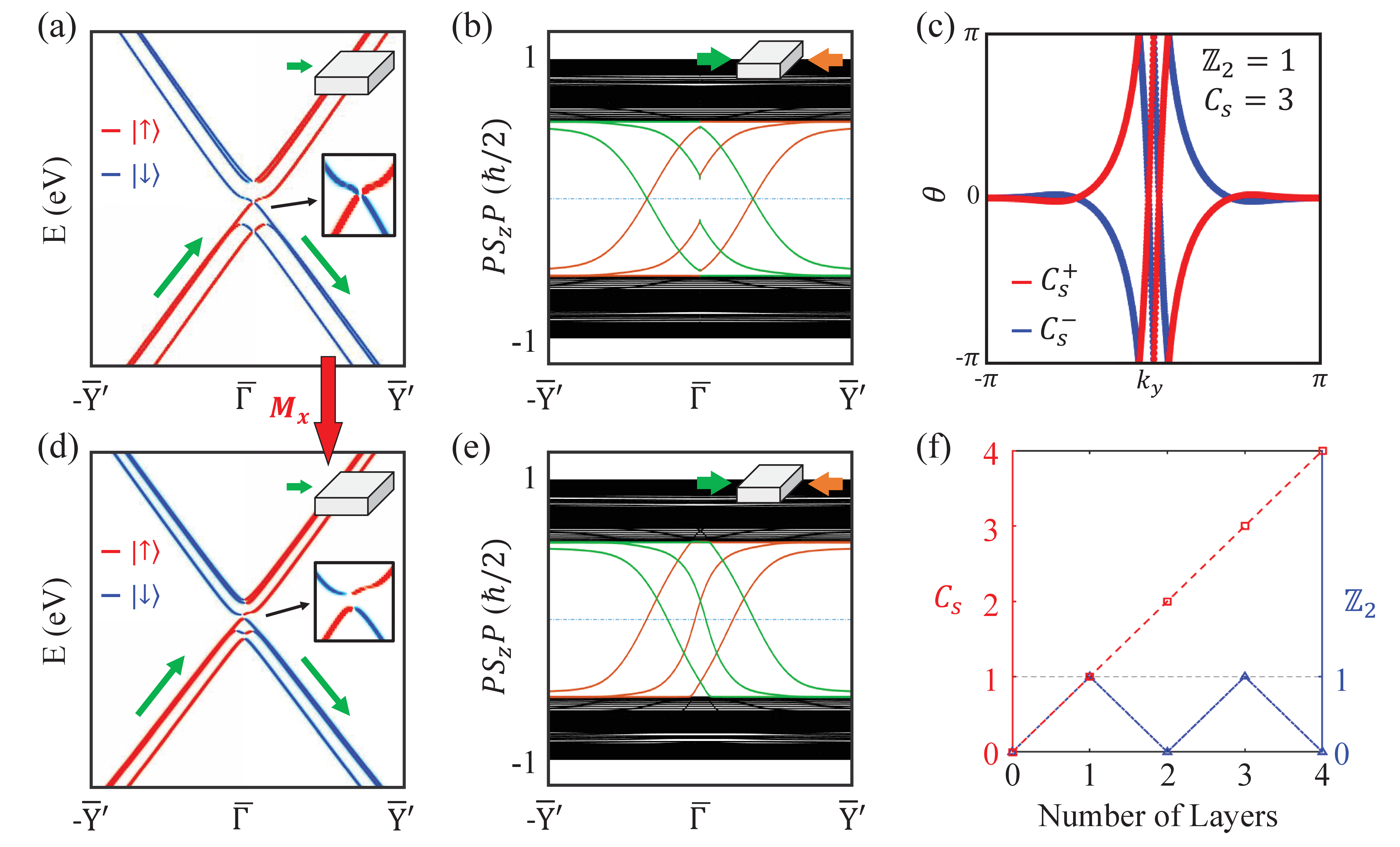}
\caption{\textbf{Trilayer Bi4Br4 and layer-dependent spin Hall conductivity.} (a)Spin-polarized edge states of trilayer $\rm{Bi_4Br_4}$ with time-reversal symmetry in the \textit{E-k} map. The green arrows indicate the direction of $\hat{S}_z$ propagation on the edge states from the LHS boundary. The green arrow in the inset denotes the edge states from the LHS boundary. (b)$\left \langle PS_zP \right \rangle$ feature edge states in trilayer $\rm{Bi_4Br_4}$ with time-reversal symmetry. The green and orange lines denote the nontrivial edge states in the edge feature spectrum from the LHS boundary and RHS boundary, respectively. The green and orange arrows in the inset indicate the edge states from the LHS boundary and RHS boundary, respectively. (c)The spin-resolve Wilson’s loop indicates the $C_s=3$ and $\mathbb{Z}_1=0$. The red and blue line indicates the $C_s^+$ and $C_s^-$, respectively. (d)Same as (a), but time-reversal symmetry breaking. The green arrows indicate the direction of $\hat{S}_z$ propagation on the edge states from LHS boundary. The green arrow in the inset indicates the edge states from the LHS boundary. (e)Same as (b), but time-reversal symmetry breaking. The green and orange lines denote the nontrivial edge states in the edge feature spectrum from the LHS boundary and RHS boundary, respectively. The green and orange arrows in the inset indicate the edge states from the LHS boundary and RHS boundary, respectively. (f)The $C_s$ and $\mathbb{Z}_2$ as a function of the number of $\rm{Bi_4Br_4}$ layers.
}
\label{fig:04}
\end{figure}

\par Proceeding to trilayer $\rm{Bi_4Br_4}$. The band structure of the edge states in trilayer $\rm{Bi_4Br_4}$ closely resembles a combination of the edge states from the monolayer and bilayer counterparts (Fig.~\ref{fig:04}(a)). One topological Dirac edge state with time-reversal symmetry preserved at $\bar{\Gamma}$ point and two attached additional gapped edge bands around the Fermi level support a 2D topological insulator phase with $\mathbb{Z}_2=1$ in trilayer $\rm{Bi_4Br_4}$. The SChN $C_s$ of trilayer $\rm{Bi_4Br_4}$ calculated from the eigenstates of the spin sector in the feature spectrum is determined to be $C_s=3$ (Fig.~\ref{fig:04}(c)). The edge feature spectrum calculation presents pairs of non-trivial edge states manifesting in the gap of bulk feature eigenvalues (Fig.~\ref{fig:04}(b)), which can be explained as the $\hat{S}_z$ spin flip on the gapped edge band structure of trilayer $\rm{Bi_4Br_4}$ in the \textit{E-k} map (green arrows in Fig.~\ref{fig:04}(a)). Our calculations display the total number of non-trivial edge states in trilayer $\rm{Bi_4Br_4}$ as three, one non-trivial state from the \textit{E-k} map and two non-trivial states from the feature spectrum, which is equivalent to the SChN $C_s=3$. Most importantly, this is an unanimous signature of feature-energy duality. As the Zeeman exchange field $M_x$ is involved, the topological Dirac edge state in the \textit{E-k} map opens a gap due to time-reversal symmetry breaking (Fig.~\ref{fig:04}(d)).  Concurrently, the third non-trivial edge state around $\bar{\Gamma}$ point emerges in the edge feature spectrum (Fig.~\ref{fig:04}(e)). These instances not only illustrate the feature-energy duality in realistic material but also demonstrate that the bulk-boundary correspondence remains universally valid based on the combination of the non-trivial states in the \textit{E-k} map and the feature spectrum (Fig.~\ref{fig:05}).

\par Fig.~\ref{fig:04}(f) presents the topological invariant numbers, $C_s$ and $\mathbb{Z}_2$, as functions of the stacking layers of $\rm{Bi_4Br_4}$. Notably, the SChN $C_s$ exhibits linear increases with the number of layers, while $\mathbb{Z}_2$ index oscillates between 1 and 0 for odd and even layers, respectively. This phenomenon arises from the requirement of symmetry protection for the existence of gapless edge states in $\mathbb{Z}_2$ topological insulators, such as time-reversal or crystalline symmetry. In an even number of layers, the interlayer coupling induces the wavefunction hybridization of the topological edge states originating from each layer. In the lacking of additional crystal symmetry protection, this hybridization effect opens a gap in the \textit{E-k} map, resulting in $\mathbb{Z}_2=0$. In contrast, the SChN $C_s$ exhibits feature-energy duality, independent of symmetry protection, whose value is equal to the total number of gapless edge states in the \textit{E-k} map and non-trivial edge states in the $\hat{S}_z$ feature spectrum. In simpler terms, the emergence of non-trivial edge states of QSHI in the \textit{E-k} map is not a prerequisite unless the system also qualifies as a $\mathbb{Z}_2$ topological insulator.

\section{\label{sec:level1}Discussion and Conclusion}

\par
\begin{figure}[ht]
\centering
\includegraphics[width=\linewidth]{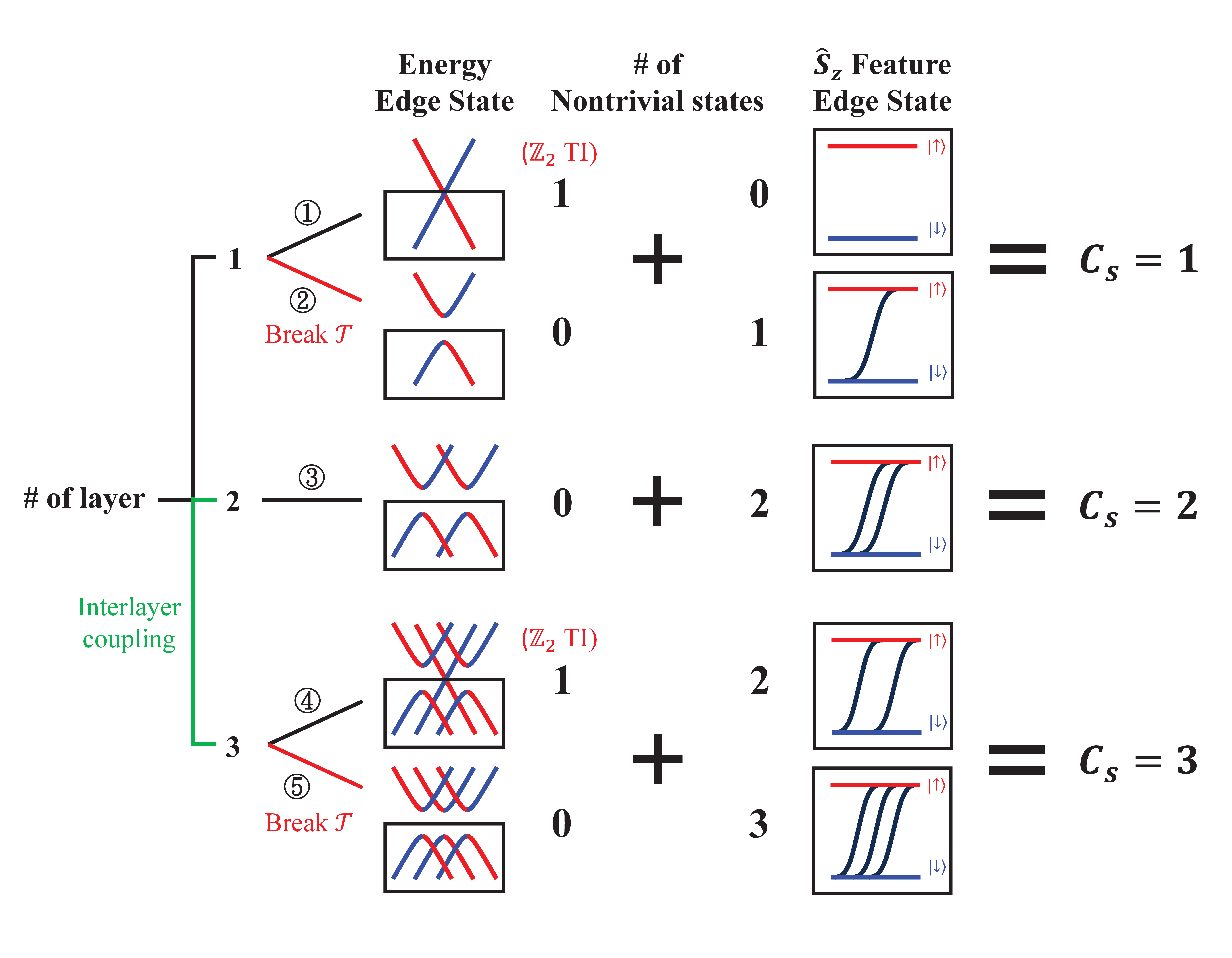}
\caption{\textbf{The feature-energy duality.} (a)These examples illustrate that the total number of gapless edge states in the \textit{E-k} map and nontrivial edge states in the $\hat{S}_z$ feature spectrum equals the spin Chern number $C_s$ of multilayer quantum spin Hall insulator.
}
\label{fig:05}
\end{figure}

 We discuss the universality of the feature-energy duality and the bulk-boundary correspondence in multilayer QSHI, as depicted in Fig.~\ref{fig:05}. In monolayer QSHI, the gapless edge state emerges in the \textit{E-k} map, while the edge feature spectrum exhibits a gap. Corresponding, topological invariant numbers are $C_s=1$ and $\mathbb{Z}_2=1$. In the second scenario, the edge state opens a gap due to time-reversal symmetry breaking. A non-trivial edge state appears in the edge feature spectrum, connecting the spin-up and spin-down sectors of bulk feature eigenvalues. Moving on to bilayer (the third case in Fig.~\ref{fig:05}), the edge states open a gap in the \textit{E-k} map due to interlayer coupling. Conversely, a pair of non-trivial edge states emerge in the edge feature spectrum. Consequently, the topological invariant numbers of bilayer QSHI are $C_s=2$ and $\mathbb{Z}_2=0$. The last two cases exhibit the same topological invariant numbers ($C_s=3$ and $\mathbb{Z}_2=1$), yet their edge state behaviors differ significantly. In the fourth case, one gapless edge state and two attached gapped edge bands appear in the \textit{E-k} map, while two non-trivial edge states manifest in the edge feature spectrum. The edge states open a gap in the \textit{E-k} map result from the time-reversal symmetry breaking perturbation in the fifth case. Simultaneously, the third non-trivial edge state emerges in the edge feature spectrum. These examples collectively demonstrate that the SChN $C_s$ equals the total number of gapless edge states in the \textit{E-k} map and non-trivial edge states in the feature spectrum.

\par Our investigation reveals that the SChN $C_s$ can always be well-defined through the feature spectrum topology \cite{NO10,NO12}, even when the $\hat{S}_z$ is non-conserved under gap-opening perturbations such as Rashba SOC and Zeeman exchange field in the Hamiltonian. As a result, we speculate that the topological phases classified by the feature operator may extend beyond conventional symmetry constraints. This concept aligns with the quantum anomalous Hall effect, where the Chern number does not necessitate additional symmetry protection. The feature spectrum topology offers a robust perspective that broadens the classification of the topological phase to a more fundamental level, transcending the limitations of symmetry-based considerations.

\par In conclusion, our exploration of bulk-boundary correspondence through feature spectrum topology, employing the $\hat{S}_z$ feature operator in both monolayer and bilayer Kane-Mele models, has revealed a feature-energy duality. Our calculation unveils that the combination of gapless edge states in the \textit{E-k} map and non-trivial edge states in the feature spectrum equals the SChN in multilayer QSHI. Additionally, we propose the van der Waals material $\rm{Bi_4Br_4}$ as a candidate. Our results validate the feature-energy duality and further identify Bi4Br4 as a layer-dependent high-spin Chern insulator. The discovery of the feature spectrum topology not only deepens our comprehension of the classification of topological phases but also opens avenues for the exploration of QSHI with high-spin Chern number.

\section{\label{sec:level1}Method}

\par The first-principle calculations on the $\rm{Bi_4Br_4}$ thin film were performed using the Vienna \textit{ab Initio} simulation package (VASP) \cite{NO28,NO29,NO30} with the projector augmented wave approach and the generalized-gradient approximation (GGA) \cite{NO31}. The spin-orbit coupling (SOC) was included self-consistently in the calculations of electronic structures with a Gamma-centered mesh 11×11×1. The experimental lattice parameters of $\rm{Bi_4Br_4}$ were used, with a=6.998$\rm{\AA}$, b=4.338$\rm{\AA}$, and 71.945$^{\rm{\circ}}$ angle between a-b axis. The interlayer distance between each layer is 2.456$\rm{\AA}$ in multilayer systems. To calculate the feature spectrum, edge state, spin Chern number, and spin Hall conductivity, we constructed the tight-binding Hamiltonian, where the matrix elements were calculated by projecting onto Wannier orbitals, which used the VASP2WANNIER90 interface \cite{NO32,NO33,NO34}. The Bi $p$ and Br $p$ orbitals were used without performing the procedure for maximizing localization.

\section*{Acknowledgement}
T.-R.C. was supported by the 2030 Cross-Generation Young
Scholars Program from the National Science and Technology Council (NSTC) in Taiwan (Program No.
MOST111-2628-M-006-003-MY3), National Cheng Kung University (NCKU), Taiwan, and National Center
for Theoretical Sciences, Taiwan. This research was supported, in part, by Higher Education Sprout Project,
Ministry of Education to the Headquarters of University Advancement at NCKU. The work at Northeastern University was supported by the National Science Foundation through NSF-ExpandQISE award \#2329067 and it benefited from Northeastern University’s Advanced Scientific Computation Center and the Discovery Cluster.


%

\end{document}